\documentstyle[12pt]{article}

\global\arraycolsep=2pt

\begin{document}

\begin{titlepage}

\begin{flushright}
CERN-TH.7070/93\\
November 1993
\end{flushright}

\vspace{0.3cm}

\begin{center}
\Large\bf A Virial Theorem for the Kinetic Energy\\
of a Heavy Quark inside Hadrons
\end{center}

\vspace{0.8cm}

\begin{center}
Matthias Neubert\\
{\sl Theory Division, CERN, CH-1211 Geneva 23, Switzerland}
\end{center}

\vspace{0.8cm}

\begin{abstract}
The formalism of the heavy quark effective theory is used to derive
the field-theory analog of the virial theorem, which relates the
matrix element of the kinetic energy of a heavy quark inside a hadron
to a matrix element of the gluon field strength tensor. The existing
QCD sum rule calculations of the kinetic energy are not consistent
with this theorem.
\end{abstract}

\centerline{(submitted to Physics Letters B)}

\end{titlepage}

\section{Introduction}

The development of the heavy quark effective theory (HQET) has led to
much progress in the theoretical understanding of the properties of
hadrons, in which a heavy quark $Q$ interacts with light degrees of
freedom predominantly by the exchange of soft gluons
\cite{Eich,Geor,Mann,Falk,habil}. It allows one to construct a
systematic expansion of hadronic parameters (such as masses, decay
constants, or transition form factors) in powers of $1/m_Q$. To
leading order, the effective theory is explicitly invariant under a
spin-flavor symmetry, which relates states containing a heavy quark
of different spin or flavor, but with the same velocity
\cite{Volo,Isgu}. This symmetry imposes important constraints on the
hadronic matrix elements of heavy quark current operators, which play
a central role in the description of both exclusive and inclusive
weak decays of heavy mesons and baryons. For example, in the heavy
quark limit the normalization of the amplitude for the exclusive
semileptonic decay $\bar B\to D^*\ell\,\bar\nu$ can be predicted at
the kinematic point where the two heavy mesons have the same
velocity, and there are no corrections of first order in $1/m_c$ or
$1/m_b$ \cite{Luke}. This process is thus ideal for a reliable
determination of the Kobayashi-Maskawa matrix element $V_{cb}$
\cite{Vcb}. For inclusive semileptonic decays of $B$ mesons, one can
prove that to leading order in $1/m_b$ the decay rates and spectra
agree with the parton model prediction, and the leading
nonperturbative corrections are of order $1/m_b^2$
\cite{Chay,Bigi,Blok,MaWe,Thom}.

In HQET, a heavy quark bound inside a hadron moving with
four-velocity $v$ is represented by a velocity-dependent field
$h_v(x)$, which is related to the conventional spinor field $Q(x)$ by
a phase redefinition \cite{Geor}. When the total momentum is written
as $p_Q=m_Q\,v+k$, the field $h_v$ carries the residual momentum $k$,
which results from soft interactions of the heavy quark with light
degrees of freedom and is typically of order $\Lambda_{\rm QCD}$.
Moreover, the heavy quark field in the effective theory is subject to
the constraint $\rlap/v\,h_v=h_v$, which projects out the heavy quark
components of the spinor. The antiquark components are integrated out
to obtain the effective Lagrangian \cite{Eich,Geor,Mann}
\begin{eqnarray}\label{Leff}
   {\cal{L}}_{\rm HQET} &=& \bar h_v\,i v\!\cdot\!D\,h_v
    + {1\over 2 m_Q}\,\bar h_v\,(i D_\perp)^2 h_v \nonumber\\
   &&+ C(m_Q)\,{g_s\over 4 m_Q}\,\bar h_v\,\sigma_{\mu\nu}
    G^{\mu\nu} h_v + {\cal{O}}(1/m_Q^2) \,,
\end{eqnarray}
where $D^\mu = \partial^\mu - i g_s A^\mu$ is the covariant
derivative, $D_\perp^\mu=D^\mu-v^\mu\,v\cdot D$ contains its
components perpendicular to the hadron velocity, and $G^{\mu\nu}$ is
the gluon field strength tensor defined by $[iD^\mu,iD^\nu]=i g_s
G^{\mu\nu}$. The physical origin of the operators appearing at order
$1/m_Q$ becomes transparent in the hadron's rest frame. There $(i
D_\perp)^2=\vec D^2$, and the first operator corresponds to the
kinetic energy resulting from the residual motion of the heavy quark.
The second operator describes the interaction of the heavy quark spin
with the chromo-magnetic gluon field. The coefficient $C(m_Q)$ of the
chromo-magnetic operator results from renormalization-group effects
and contains a logarithmic dependence on the heavy quark mass
\cite{FGL}. The coefficient of the kinetic operator is not
renormalized \cite{LuMa}.

In order to construct a systematic $1/m_Q$ expansion, one works with
the eigenstates of the leading-order term in the effective Lagrangian
(\ref{Leff}), supplemented by the standard QCD Lagrangian for the
light quarks and gluons, and treats the $1/m_Q$ corrections as
perturbations. The states of HQET are thus different from the
physical states of the full theory. They correspond to the would-be
hadrons composed of an infinitely heavy quark and light degrees of
freedom. Because of the spin symmetry of the effective theory, states
which are related to each other by a spin flip of the heavy quark are
degenerate in mass. For instance, the ground-state pseudoscalar ($P$)
and vector ($V$) mesons form a doublet under the spin symmetry. Only
beyond the leading order in $1/m_Q$, a small mass splitting is
induced by the presence of the spin-symmetry breaking chromo-magnetic
interaction in (\ref{Leff}). In fact, the masses of the physical
states obey an expansion
 \begin{equation}
   M = m_Q + \bar\Lambda - {\lambda_1\over 2 m_Q}
   - d_M\,{\lambda_2(m_Q)\over 2 m_Q} + {\cal{O}}(1/m_Q^2) \,,
\end{equation}
where $d_P=3$ for a pseudoscalar meson, and $d_V=-1$ for a vector
meson. The parameter $\bar\Lambda$ corresponds to the effective mass
of the light degrees of freedom in the $m_Q\to\infty$ limit
\cite{AMM}, and the coefficients $\lambda_1$ and $\lambda_2(m_Q)$
parameterize the contributions from the kinetic and the
chromo-magnetic operator \cite{FaNe}. They are defined as ($M=P$ or
$V$)
\begin{eqnarray}\label{lamdef}
   {\langle M(v)|\,\bar h_v\,(i D_\perp)^2 h_v\,| M(v)\rangle
    \over \langle M(v)|\,\bar h_v\,h_v\,| M(v)\rangle}
   &=& \lambda_1 = -2 m_Q\,K_Q \,, \nonumber\\
   && \\
   C(m_Q)\,{\langle M(v)|\,\bar h_v\,g_s\,\sigma_{\mu\nu} G^{\mu\nu}
    h_v\,| M(v)\rangle \over\langle M(v)|\,\bar h_v\,h_v\,|
    M(v)\rangle}
   &=& 2 d_M\,\lambda_2(m_Q) \,. \nonumber
\end{eqnarray}
The quantity $\lambda_1$ is a fundamental parameter of HQET, which is
independent of the heavy quark mass. We note that $K_Q=-\lambda_1/2
m_Q$ is the expectation value (in the effective theory) of the
kinetic energy of the heavy quark in the hadron's rest frame. It is
the same for pseudoscalar and vector mesons. The second parameter,
$\lambda_2$, contains a logarithmic dependence on $m_Q$ arising from
the short-distance coefficient $C(m_Q)$.

Both $\lambda_1$ and $\lambda_2$ play a crucial role in the
description of power corrections to the heavy quark limit. For
instance, the leading nonperturbative corrections to inclusive
semileptonic decay rates are completely determined in terms of these
two quantities \cite{Bigi,Blok,MaWe,Thom}. An estimate of $\lambda_2$
can be extracted from the observed mass splitting between $B$ and
$B^*$ mesons: $\lambda_2(m_b)\simeq\frac{1}{4}(m_{B^*}^2-m_B^2)\simeq
0.12$ GeV$^2$. The parameter $\lambda_1$, however, is not directly
related to an observable quantity. In the phenomenological model of
Ref.~\cite{ACM}, the kinetic energy results from the Fermi motion of
the heavy quark inside the hadron, and one expects $-\lambda_1\approx
p_F^2\approx 0.1$ GeV$^2$, where $p_F$ is the Fermi momentum. On the
other hand, a recent QCD sum rule analysis predicts the much larger
value $-\lambda_1=0.6\pm 0.1$ GeV$^2$ \cite{BaBr}.

In this paper, we shall use the formalism of HQET to derive the
field-theory analog of the virial theorem, which relates $\lambda_1$
to a matrix element of the gluon field strength tensor between hadron
states of {\sl different\/} velocity. We show that the existing QCD
sum rule calculations of $\lambda_1$ in Refs.~\cite{BaBr,Subl,Elet}
are not consistent with this general relation. The results obtained
from these analyses should therefore be taken with some caution. In
fact, we argue that the kinetic energy is probably smaller than
predicted in these approaches.

\section{Derivation of the Virial Theorem}

In order to derive the virial theorem, one has to study in detail the
structure of hadronic matrix elements of local dimension-five current
operators in HQET. Such an analysis was performed in
Ref.~\cite{FaNe}, where the second-order power corrections to meson
and baryon weak decay form factors have been investigated. Here we
shall elaborate on some results obtained in this work.

Since the states of HQET are taken to be the eigenstates of the
leading term in the effective Lagrangian (\ref{Leff}), matrix element
evaluated between these states have well-defined transformation
properties under the spin-flavor symmetry group. In addition, these
matrix elements are constrained by the equations of motion of HQET,
and by the requirement of Lorentz covariance. These constraints are
most elegantly incorporated in the covariant tensor formalism
introduced in Ref.~\cite{Falk}. Hadrons containing a heavy quark and
light degrees of freedom are represented by tensors with the correct
transformation properties under the Lorentz group and under rotations
of the heavy quark spin. For instance, the doublet of the
ground-state pseudoscalar and vector mesons is represented by
$4\times 4$ Dirac matrices
\begin{equation}
   {\cal{M}}(v) = {1+\rlap/v\over 2\sqrt{2}}\,
   \cases{
    - \gamma_5 &; pseudoscalar meson $P(v)$, \cr
    \rlap/\epsilon &; vector meson $V(v,\epsilon)$, \cr}
\end{equation}
where $\epsilon^\mu$ is the polarizartion vector of the vector meson
($\epsilon\cdot v=0$). We use a nonrelativistic normalization of
states, such that ($M=P$ or $V$)
\begin{equation}
   \langle M(v)|\,\bar h_v\,h_v\,|M(v)\rangle
   = -{\rm Tr}\Big\{\, \overline{\cal{M}}(v)\,{\cal{M}}(v) \,\Big\}
   = 1 \,.
\end{equation}
As in this example, meson matrix elements of operators in HQET can
always be written as traces over these wave functions; however, for
more complicated matrix elements one has to include appropriate
tensors for the light degrees of freedom, too.

Consider, as an example, the matrix element of a dimension-five
current operator containing the gluon field strength tensor between
meson states with different velocity. Using the tensor formalism, one
can write
\begin{equation}\label{phidef}
   \langle M(v')|\,\bar h_{v'}\Gamma\,i g_s G^{\mu\nu} h_v\,
   |M(v)\rangle = -{\rm Tr}\Big\{\,\phi^{\mu\nu}(v,v')\,
   \overline{\cal{M}}(v')\,\Gamma {\cal{M}}(v) \,\Big\} \,
\end{equation}
where $\Gamma$ may be an arbitrary combination of Dirac matrices. The
object $\phi^{\mu\nu}(v,v')$ represents the light degrees of freedom.
It is a most complicated hadronic quantity which, however, can only
depend on the meson velocities. Using the projection property
$\rlap/v\,{\cal{M}}(v) = {\cal{M}}(v) = -{\cal{M}}(v)\,\rlap/v$ of
the meson wave functions, one finds that the most general
decomposition of the form factor is
\begin{eqnarray}\label{phiidef}
   \phi^{\mu\nu}(v,v') &=& \phi_1(w)\,(v^\mu v'^\nu - v^\nu v'^\mu)
    \nonumber\\ \phantom{ \bigg[ }
   &+& \phi_2(w)\,\Big[ (v-v')^\mu \gamma^\nu - (v-v')^\nu \gamma^\mu
    \Big] + i\phi_3(w)\,\sigma^{\mu\nu} \,,
\end{eqnarray}
where $w=v\cdot v'$. $T$-invariance of the strong interactions
requires that the scalar functions $\phi_i(w)$ be real. Note that the
matrix element must be invariant under complex conjugation
accompanied by an interchange of the indices and the velocity (and
polarization) vectors. This forbids a term proportional to
$\big[(v+v')^\mu\gamma^\nu-(v+v')^\nu\gamma^\mu\big]$.

Using the equations of motion of HQET, one can show that the
normalization of the functions $\phi_i(w)$ at zero recoil ($w=1$) is
related to the parameters $\lambda_1$ and $\lambda_2$. One finds
\cite{FaNe}
\begin{equation}\label{phirel}
   \phi_1(1) - \phi_2(1) - {1\over 2}\,\phi_3(1)
   = -{\lambda_1\over 3} \,, \qquad \phi_3(1) = \lambda_2 \,.
\end{equation}
For completeness, we give the details of the derivation of this
result in the appendix. Not that the first equation relates
$\lambda_1$ to form factors parameterizing the matrix elements of the
operator containing the gluon field strength tensor.

In order to proceed, we evaluate (\ref{phidef}) for $\Gamma=1$, both
for pseudoscalar and vector mesons. In the latter case the
polarization vector is chosen to be the same in the initial and final
state. We obtain
\begin{eqnarray}
   &&\langle P(v')|\,\bar h_{v'}\,i g_s G^{\mu\nu} h_v\,|P(v)\rangle
    = \langle V(v',\epsilon)|\,\bar h_{v'}\,i g_s G^{\mu\nu} h_v\,
    |V(v,\epsilon)\rangle \nonumber\\ \phantom{ \Bigg[ }
   &&\quad = {1\over 2}\,(v^\mu v'^\nu - v^\nu v'^\mu)\,
    \Big[ (w+1)\phi_1(w) - 2\phi_2(w) - \phi_3(w) \Big] \nonumber\\
   &&\quad = (v^\mu v'^\nu - v^\nu v'^\mu)\,\bigg[
    -{\lambda_1\over 3} + {\cal{O}}(w-1) \bigg] \,.
\end{eqnarray}
A corresponding relation holds for the ground-state $\Lambda_Q$
baryons containing a heavy quark, too. In this case, the analogs of
the functions $\phi_2(w)$ and $\phi_3(w)$ vanish, and the analog of
$\phi_1(w)$ is normalized at zero recoil to $-\tilde\lambda_1/3$
\cite{FaNe}. The parameter $\tilde\lambda_1$ parameterizes the baryon
matrix element of the kinetic operator and is defined in analogy to
(\ref{lamdef}). Taking into account that both $\lambda_1$ and
$\tilde\lambda_1$ are proportional to the kinetic energy $K_Q$ of the
heavy quark inside the hadron, we obtain the virial theorem
\begin{equation}\label{virial}
   {\langle H_Q(v')|\,\bar h_{v'}\,i g_s G^{\mu\nu} h_v\,
    |H_Q(v)\rangle\over
    \langle H_Q(v)|\,\bar h_v\,h_v\,|H_Q(v)\rangle}
   = {2 m_Q\over 3}\,(v^\mu v'^\nu - v^\nu v'^\mu)\,
   \Big[ K_Q + {\cal{O}}(w-1) \Big] \,,
\end{equation}
which in this form is independent of the normalization of the states.
The hadron $H_Q$ can be any of the ground-state heavy mesons or
baryons.

It is instructive to evaluate this relation in the rest frame of the
initial hadron, where the nonvanishing components of the tensor on
the right-hand side are those with $\mu=0$ or $\nu=0$. We find
\begin{equation}\label{vir2}
   {\langle H_Q(\vec v')|\,\bar h_{\vec v'}\,i g_s\vec E_c\,
    h_{\vec 0}\,|H_Q(\vec 0)\rangle\over\langle H_Q(\vec 0)|\,
    \bar h_{\vec 0}\,h_{\vec 0}\,|H_Q(\vec 0)\rangle}
    = - {2 m_Q\over 3}\,\vec v'\,K_Q + {\cal{O}}(\vec v'^3) \,,
\end{equation}
where $E_c^i=-G^{0i}$ are the components of the chromo-electric
field.

In quantum mechanics, the matrix element on the left-hand side can
be represented as
\begin{equation}
   \langle H_Q|\,\exp(i m_H\,\vec v'\cdot{\bf\vec x})\,
   i g_s\vec E_c({\bf\vec x})\,|H_Q\rangle \,,
\end{equation}
where the states are at rest and nomalized to unity. The final state
hadron with momentum $m_H\,\vec v'$ has been related to a hadron at rest by
a boost operator, where
${\bf\vec x}$ denotes the operator for the center-of-mass coordinate.
In the heavy quark limit, one can identify ${\bf\vec x}$ with the
position operator for the heavy quark. We can now expand the
exponential in powers of $\vec v'$, using that by rotational
invariance
\begin{eqnarray}
   \langle H_Q|\,\vec E_c({\bf\vec x})\,|H_Q\rangle &=& 0 \,,
    \nonumber\\ \phantom{ \Bigg[ }
   \langle H_Q|\,{\bf x}^i\,E_c^j({\bf\vec x})\,|H_Q\rangle
   &=& {\delta^{ij}\over 3}\,\langle H_Q|\,
    {\bf\vec x}\cdot\vec E_c({\bf\vec x})\,|H_Q\rangle \,.
\end{eqnarray}
Equating the terms linear in $\vec v'$ in (\ref{vir2}), and using
that $m_H/m_Q=1$ to leading order in $1/m_Q$, we obtain
\begin{equation}
   2 K_Q = \langle H_Q|\,g_s\,
   {\bf\vec x}\cdot\vec E_c({\bf\vec x})\,|H_Q\rangle \,.
\end{equation}
In the abelian case, where the chromo-electric field can be written
in terms of a potential, and $g_s \vec E_c(\vec x)=\vec\nabla\,
V(\vec x)$ is the gradient of the potential energy of the heavy quark
interacting with the background field of the light degrees of
freedom, this is nothing but the classical virial theorem.

\section{Conclusions}

Using the formalism of the heavy quark effective theory, we have
derived the field-theory version of the virial theorem, which relates
the matrix element of the kinetic energy of a heavy quark inside a
hadron to a matrix element of the gluon field strength tensor between
hadron states with different velocity. This theorem is interesting in
that it shows that the two parameters $\lambda_1$ and $\lambda_2$,
which play an important role in the description of power corrections
to the heavy quark limit, have a similar origin.

The virial theorem has direct implications for existing QCD sum rule
calculations of the kinetic energy. In these analyses, the main
contribution comes from a diagram not containing gluons (the bare
quark loop). However, when the kinetic energy is calculated by
constructing a sum rule for the left-hand side of (\ref{virial}),
such a graph does not appear. Hence, it cannot contribute to the sum
rule for $\lambda_1$. In other words, the virial theorem makes
explicit an ``intrinsic smallness'' of the kinetic energy, which in
the existing sum rule calculations would have to result from a
cancellation of several large contributions. We conclude that the
numerical results for the kinetic energy quoted in
Refs.~\cite{BaBr,Subl,Elet} should be taken with caution. A new QCD
sum rule analysis, which incorporates the virial theorem, will be
presented elsewhere \cite{future}. Since in this sum rule the leading
contribution comes from a two-loop diagram, the resulting value of
the kinetic energy will be considerably smaller. \bigskip

{\it Acknowledgments:\/}
It is a pleasure to thank Thomas Mannel for useful discussions, and
Adam Falk for collaboration on subjects closely related to this work.
\bigskip\bigskip\bigskip\bigskip

{\Large\bf Appendix}
\renewcommand{\theequation}{A.\arabic{equation}}
\setcounter{equation}{0}
\bigskip

For completeness, we give below the main steps of a derivation
presented in Ref.~\cite{FaNe}, which leads to the zero recoil
conditions (\ref{phirel}). Consider the matrix element
\begin{equation}\label{psidef}
   \langle M(v')|\,(-i D^\mu\bar h_{v'})\,\Gamma\,i D^\nu h_v\,
   |M(v)\rangle = -{\rm Tr}\Big\{\,\psi^{\mu\nu}(v,v')\,
   \overline{\cal{M}}(v')\,\Gamma{\cal{M}}(v) \,\Big\} \,.
\end{equation}

The most general decomposition of the tensor form factor
$\psi^{\mu\nu}(v,v')$ involves ten invariant functions. However,
considering the complex conjugate of the above matrix element, one
finds the symmetry relation $\overline{\psi}^{\mu\nu}(v',v) =
\psi^{\mu\nu}(v,v')$, which reduces the number of invariant functions
to seven. It is convenient to perform a decomposition into symmetric
and antisymmetric parts, $\psi^{\mu\nu} = \frac{1}{2}(\psi_S^{\mu\nu}
+ \psi_A^{\mu\nu})$, and to define
\begin{eqnarray}
   \psi_S^{\mu\nu}(v,v') &=& \psi_1^S(w)\,g^{\mu\nu}
    + \psi_2^S(w)\,(v+v')^\mu (v+v')^\nu \nonumber\\
   \phantom{ \bigg[ }
   &+& \psi_3^S(w)\,(v-v')^\mu (v-v')^\nu + \psi_4^S(w)\,\Big[
    (v+v')^\mu \gamma^\nu + (v+v')^\mu \gamma^\nu \Big] \,,
    \nonumber\\
   && \\
   \psi_A^{\mu\nu}(v,v')
   &=& \psi_1^A(w)\,(v^\mu v'^\nu - v'^\mu v^\nu) \nonumber\\
   \phantom{ \bigg[ }
   &+& \psi_2^A(w)\,\Big[ (v-v')^\mu \gamma^\nu - (v-v')^\nu
    \gamma^\mu \Big] + i \psi_3^A(w)\,\sigma^{\mu\nu} \,,
   \nonumber
\end{eqnarray}
where $w=v\cdot v'$. Because of $T$-invariance of the strong
interactions, the invariant functions are real.

The equations of motion of HQET imply that, under the trace in
(\ref{psidef}), $v_\nu\,\psi^{\mu\nu}(v,v')\,\hat =\,0$. This leads
three relations among the seven functions, which can be used to
eliminate $\psi_1^S(w)$, $\psi_2^S(w)$, and $\psi_4^S(w)$.

An integration by parts relates (\ref{psidef}) to a matrix element of
an operator containing two derivatives acting on the same heavy quark
field. The result is
\begin{eqnarray}\label{partint}
   \langle M(v')|\,\bar h_{v'}\,\Gamma\,i D^\mu i D^\nu h_v\,
   |M(v)\rangle &=& \langle M(v')|\,(-i D^\mu\bar h_{v'})\,
    \Gamma\,i D^\nu h_v\,|M(v)\rangle \nonumber\\
   \phantom{ \bigg[ }
   &+& \bar\Lambda\,(v-v')^\mu\,\langle M(v')|\,
    \bar h_{v'}\,\Gamma\,i D^\nu h_v\,|M(v)\rangle \,, \nonumber\\
\end{eqnarray}
where $\bar\Lambda=M-m_Q$. The second matrix element on the
right-hand side can be written as \cite{Luke}
\begin{equation}
   \langle M(v')|\,\bar h_{v'}\,\Gamma\,i D^\nu h_v\,|M(v)\rangle
   = -{\rm Tr}\Big\{\, \xi^\nu(v,v')\,\overline{\cal{M}}(v')\,
   \Gamma\,{\cal{M}}(v) \,\Big\} \,,
\end{equation}
where
\begin{equation}
   (w+1)\,\xi^\nu(v,v') = (w\,v^\nu-v'^\nu)\,\bar\Lambda\,\xi(w)
   - \Big[ (v+v')^\nu + (w+1)\,\gamma^\nu \Big]\,\xi_3(w) \,.
\end{equation}

One can use the above results to relate the functions $\phi_i(w)$
defined in (\ref{phiidef}), as well as the function $\phi_0(w)$
defined as
\begin{equation}
   \langle M(v')|\,\bar h_{v'}\,\Gamma\,(i D_\perp)^2 h_v\,
   |M(v)\rangle = - \phi_0(w)\,{\rm Tr}\Big\{\,
   \overline{\cal{M}}(v')\,\Gamma\,{\cal{M}}(v) \,\Big\} \,,
\end{equation}
to the four functions $\psi_3^S(w)$ and $\psi_i^A(w)$. The result is
\begin{eqnarray}
   \phi_0(w) &=& -{2w+1\over w+1}\,\Big[ (w+1)\,\psi_1^A(w)
    - 2\psi_2^A(w) - \psi_3^A(w) \Big] \nonumber\\
   \phantom{ \bigg[ }
   &&\mbox{}+ 2(w-1)\,\psi_3^S(w) - \bar\Lambda^2 (w-1)\,\xi(w)
    \,, \nonumber\\ \phantom{ \bigg[ }
   \phi_1(w) &=& \psi_1^A(w) + {1\over w+1}\,\Big[
    \bar\Lambda^2 (w-1)\,\xi(w) - 2\bar\Lambda\,\xi_3(w) \Big]
    \,,\nonumber\\ \phantom{ \bigg[ }
   \phi_2(w) &=& \psi_2^A(w) - \bar\Lambda\,\xi_3(w) \,,
    \nonumber\\ \phantom{ \bigg[ }
   \phi_3(w) &=& \psi_3^A(w) \,.
\end{eqnarray}
{}From a comparison with (\ref{lamdef}), one concludes that
$\phi_0(1)=\lambda_1$ and $\phi_3(1)=\lambda_2$. This leads to the
relations given in (\ref{phirel}).


\begin{thebibliography}{999}

\bibitem {Eich}
E. Eichten and B. Hill, Phys.\ Lett.\ B {\bf 234}, 511 (1990);
{\bf 243}, 427 (1990).

\bibitem {Geor}
H. Georgi, Phys.\ Lett.\ B {\bf 240}, 447 (1990).

\bibitem {Mann}
T. Mannel, W. Roberts, and Z. Ryzak, Nucl.\ Phys.\ {\bf B368}, 204
(1992).

\bibitem {Falk}
A.F. Falk, H. Georgi, B. Grinstein, and M.B. Wise, Nucl.\ Phys.\
{\bf B343}, 1 (1990).

\bibitem {habil}
For a comprehensive review, see: M. Neubert, SLAC preprint
SLAC--PUB--6263 (1993), to appear in Physics Reports.

\bibitem {Volo}
M.B. Voloshin and M.A. Shifman, Yad.\ Fiz.\ {\bf 47}, 801 (1988)
[Sov.\ J.\ Nucl.\ Phys.\ {\bf 47}, 511 (1988)].

\bibitem {Isgu}
N. Isgur and M.B. Wise, Phys.\ Lett.\ B {\bf 232}, 113 (1989);
{\bf 237}, 527 (1990).

\bibitem {Luke}
M.E. Luke, Phys.\ Lett.\ B {\bf 252}, 447 (1990).

\bibitem {Vcb}
M. Neubert, Phys.\ Lett.\ B {\bf 264}, 455 (1991).

\bibitem {Chay}
J. Chay, H. Georgi, and B. Grinstein, Phys.\ Lett.\ B {\bf 247}, 399
(1990).

\bibitem {Bigi}
I.I. Bigi, M. Shifman, N.G. Uraltsev, and A. Vainshtein, Phys.\ Rev.\
Lett.\ {\bf 71}, 496 (1993); I.I. Bigi et al., Minnesota preprint
TPI--MINN--92/67--T (1992).

\bibitem {Blok}
B. Blok, L. Koyrakh, M. Shifman, and A.I. Vainshtein, Santa Barbara
preprint NSF--ITP--93--68 (1993).

\bibitem {MaWe}
A.V. Manohar and M.B. Wise, San Diego preprint UCSD/PTH 93--14
(1993).

\bibitem {Thom}
T. Mannel, Darmstadt preprint IKDA 93/26 (1993).

\bibitem {FGL}
A.F. Falk, B. Grinstein, and M.E. Luke, Nucl.\ Phys.\ {\bf B357}, 185
(1991).

\bibitem {LuMa}
M. Luke and A.V. Manohar, Phys.\ Lett.\ B {\bf 286}, 348 (1992).

\bibitem {AMM}
A.F. Falk, M. Neubert, and M. Luke, Nucl.\ Phys. {\bf B388}, 363 (1992).

\bibitem {FaNe}
A.F. Falk and M. Neubert, Phys.\ Rev.\ D {\bf 47}, 2965 (1993);
{\bf 47}, 2982 (1993).

\bibitem {ACM}
G. Altarelli et al., Nucl.\ Phys.\ {\bf B208}, 365 (1982).

\bibitem {BaBr}
P. Ball and V.M. Braun, M\"unchen preprint TUM--T31--42/93 (1993).

\bibitem {Subl}
M. Neubert, Phys.\ Rev.\ D {\bf 46}, 1076 (1992).

\bibitem {Elet}
V. Eletsky and E. Shuryak, Phys.\ Lett.\ B {\bf 276}, 191 (1992).

\bibitem {future}
M. Neubert, in preparation.

\end{thebibliography}
\end{document}